\documentclass[runningheads]{llncs}
\usepackage{graphicx}
\usepackage{amsmath}
\usepackage{cite}
\usepackage{subfigure}
\usepackage{multirow}
\usepackage{authblk}
\usepackage{indentfirst}
\usepackage{algorithm}
\usepackage{algorithmicx}
\usepackage{algpseudocode}
\usepackage{bm}
\usepackage{CJK}
\usepackage{float}
\usepackage{colortbl}
\usepackage{textcomp,booktabs}
\usepackage{array}
\usepackage{xcolor}%
\usepackage{amsfonts}%
\usepackage{amssymb}
\usepackage{graphicx}
\newcommand{\eqnb}{\begin{equation}}
\newcommand{\eqne}{\end{equation}}

\newtheorem{Rem}{Remark}
\begin{document}
\title{Sensitivity-Based Optimization for\\Blockchain Selfish Mining\thanks{Supported by the National Natural Science Foundation of China under grant No. 71932002}}
%
%\titlerunning{Abbreviated paper title}
% If the paper title is too long for the running head, you can set
% an abbreviated paper title here
%
\author{Jing-Yu Ma\inst{1} \ \ \and
Quan-Lin Li\inst{2}}
\authorrunning{Jing-Yu Ma and Quan-Lin Li}
% First names are abbreviated in the running head.
% If there are more than two authors, 'et al.' is used.

\institute{Bussiness School, Xuzhou University of Technology, Xuzhou 221018, China
\email{mjy0501@126.com}\\
\and
School of Economics and Management, Beijing University of Technology,\\ Beijing 100124, China\\
\email{liquanlin@tsinghua.edu.cn}}
\maketitle              % typeset the header of the contribution
\begin{abstract}
In this paper, we provide a novel dynamic decision method of blockchain selfish mining by applying the sensitivity-based optimization theory. Our aim is to find the optimal dynamic blockchain-pegged policy of the dishonest mining pool. To study the selfish mining attacks, two mining pools is designed by means of different competitive criterions, where the honest mining pool follows a two-block leading competitive criterion, while the dishonest mining pool follows a modification of two-block leading competitive criterion through using a blockchain-pegged policy. To find the optimal blockchain-pegged policy, we set up a policy-based continuous-time Markov process and analyze some key factors. Based on this, we discuss monotonicity and optimality of the long-run average profit with respect to the blockchain-pegged reward and prove the
structure of the optimal blockchain-pegged policy. We hope the methodology and results derived in this paper can shed light on the dynamic decision research on the selfish mining attacks of blockchain selfish mining.

\end{abstract}
\section{Introduction}

Blockchain is used to securely record a public shared ledger of Bitcoin payment transactions among Internet users
in an open P2P network. Though the security of blockchain is always regarded as the top priority, it is still threatened by some \emph{selfish mining attacks}. In the PoW blockchain, the probability that an individual miner can successfully mine a block becomes lower and lower,
as the number of joined miners increases. This greatly increases the mining risk of each
individual miner. In this situation, some miners willingly form a mining pool. Blockchain selfish mining leads to colluding miners in dishonest mining pools, one of which can obtain a revenue larger than their fair share. The existence of the selfish mining not only means that it is unfair to solve PoW puzzles but also is a severe flaw in integrity of blockchain.
%. Thus far, developing such selfish mining pools has become increasingly widespread.
%The fairness of blockchain is applied in a wide range of practical
%areas including finance, health care, smart energy, reputation
%systems, security management, Internet of Things, sharing economy, supply
%chain management and so on. Readers may refer to recent excellent survey paper, such as Li et al. \cite{Li:2019}, Li et al. \cite{Li:2018}, Jaoude and Saade \cite{Jao:2019}, Fauziah et al. \cite{Fau:2020}, Sharma et al. \cite{Sha:2020}. Although some methods are applied to ensure the fairness
%of the blockchain, there still exist selfish mining attacks.

The existence of such selfish mining attacks was first proposed by Eyal and Sirer \cite{Eya:2014}, they set up a Markov chain to express the
dynamic of the selfish mining attacks efficiently. After then, some researchers extended and generalized such a similar method to discuss other attack
strategies of blockchain. The newest work is Li et al. \cite{Li:2020}, which provided a new theoretical framework of pyramid
Markov processes to solve some open and fundamental problems of blockchain selfish mining under a rigorously mathematical setting. G\"{o}bel et al. \cite{Gob:2016}, Javier and Fralix \cite{Jav:2020} used two-dimensional Markov chain to study the selfish mining. Furthermore, some key research includes stubborn mining by Nayak et al. \cite{Nay:2016}; Ethereum by Niu and Feng \cite{Niu:2019}; multiple mining pools by Jain \cite{Jai:2019}; multi-stage blockchain by Chang et al. \cite{Chang:2019}; no block reward by Carlsten et al. \cite{Car:2016};
power adjusting by Gao et al. \cite{Gao:2019}.

In the study of blockchain selfish mining, it is a key to develop effective
optimal methods and dynamic control techniques. However, little work has been
done on applying Markov decision processes (MDPs) to set up optimal dynamic
control policies for blockchain selfish mining. In general, such a study is
more interesting, difficult and challenging. Based on Eyal and Sirer
\cite{Eya:2014},\ Sapirshtein et al.\ \cite{Sap:2016}\ extended the underlying
model for selfish mining attacks, and provided an algorithm to find $\epsilon
$-optimal policies for attackers within the model through MDPs. Furthermore,
W\"{u}st \cite{wus:2016} provided a quantitative framework based on MDPs to
analyse the security of different PoW blockchain instances with various
parameters against selfish mining. Gervais et al. \cite{Ger:2016} extended the
MDP of\ Sapirshtein et al.\ \cite{Sap:2016} to determine optimal adversarial
strategies for selfish mining. Recently, Zur et al. \cite{Zur:2020} presented a novel technique called ARR (Average Reward Ratio) MDP to
tighten the bound on the threshold for selfish mining in Ethereum.

The purpose of this paper is to apply the MDPs to set up an optimal
parameterized policy (i.e., blockchain-pegged policy) for blockchain selfish
mining. To do this, we first apply the sensitivity-based optimization theory
in the study of blockchain selfish mining, which is an effective tool proposed for performance optimization of Markov
systems by Cao \cite{Cao:2007}. Li \cite{Li:2010} and Li and Cao \cite{Li:2004} further extended and generalized such a method to a more general framework of perturbed Markov processes. A key idea in the sensitivity-based optimization theory is the
performance difference equation that can quantify the performance difference
of a Markov system under any two different policies. The
performance difference equation gives a straightforward perspective to study
the relation of the system performance between two different policies, which provides more
sensitivity information. Thus, the sensitivity-based optimization theory has
been applied to performance optimization in many practical
areas. For example, the energy-efficient data centers by Xia et al.
\cite{Xia:2018b} and Ma et al. \cite{Ma:2019a, Ma:2019b}; the inventory
rationing by Li et al. \cite{LiL:2019}; the multi-hop wireless networks by
Xia and Shihada \cite{Xia:2015} and the finance by Xia \cite{Xia:2020}.

The main contributions of this paper are twofold. The first one is to apply
the sensitivity-based optimization theory to study the blockchain selfish
mining for the first time, in which we design a modification of two-block
leading competitive criterion for the dishonest mining pool. Different from previous works in
the literature for applying an ordinary MDP to against the selfish mining
attacks, we propose and develop an easier and more convenient dynamic decision
method for the dishonest mining pool: the sensitivity-based optimization theory. Crucially, this sensitivity-based
optimization theory may open a new avenue to the optimal blockchain-pegged
policy of more general blockchain systems. The second contribution of this paper is to
characterize the optimal blockchain-pegged policy of the dishonest mining pool. We analyze the monotonicity and optimality of the long-run average profit with respect to the blockchain-pegged policies under some restrained rewards. We obtain the structure of optimal blockchain-pegged policy is
related to the blockchain reward. Therefore, the results of this paper give
new insights on understanding not only competitive criterion design of
blockchain selfish mining, but also applying the sensitivity-based
optimization theory to dynamic decision for the dishonest mining pool. We hope that the
methodology and results given in this paper can shed light on the study of
more general blockchain systems.

The remainder of this paper is organized as follows. In Section 2, we describe
a problem of blockchain selfish mining with two different mining pools. In
Section 3, we establish a policy-based continuous-time Markov process and
introduce some key factors. In Section 4, we discuss the monotonicity and optimality of the long-run average profit with
respect to the blockchain-pegged policy by the sensitivity-based optimization
theory. Finally, we give some concluding remarks in Section 5.

\section{Problem Description}

In this section, we give a problem description of blockchain selfish mining
with two different mining pools. Also, we provide system structure, operational mode and mathematical notations.

\textbf{Mining pools:} There are two different mining pools: honest and dishonest mining pools.

(a) The honest mining pool follows the Bitcoin protocol. If he mines a block, he will broadcast to whole
community immediately. To avoid the 51\% attacks, we assume the honest mining pool are majority in the blockchain system.

(b) The dishonest mining pool has the selfish mining attacks. When the
dishonest mining pool mines a block, he can earn more unfair revenue. Such revenue will attract some rational honest miners to jump into the dishonest mining pool. We denote the
efficiency-increased ratio of the dishonest mining pool and the net jumping's
mining rate by$\ \tau\ $and $\gamma$, respectively.

\textbf{Selfish mining processes: }We assume that the blocks mined by the
honest and dishonest mining pools have formed two block branches forked a tree root,
and the growths of the two block branches are two Poisson processes with
block-generating rates $\alpha_{1}$ and $\alpha_{2}$, respectively. In the
honest mining pool, the block-generating rate $\alpha_{1}$ is equal to the net
mining rate, but the situation in the dishonest mining pool is a bit
different. The block-generating rate for the dishonest mining pool is $\alpha
_{2}=\widetilde{\alpha}_{2}\left(  1+\tau\right)$, where $\widetilde{\alpha
}_{2}$ is regarded as the net mining rate when all the dishonest miners become honest. Following the protocol can not earn more
rewards, the honest miners like to jump to the dishonest mining pool with the net jumping rate $\gamma$, the real
mining rates of the honest and dishonest mining pools are given by $\lambda
_{1}=\alpha_{1}-\gamma$ and $\lambda_{2}=\left(  \widetilde{\alpha}_{2}%
+\gamma\right)  \left(  1+\tau\right)  $, respectively.

Note that mining costs of both mining pools contains two parts: (a) Power
consumption cost. Let $c_{P}$ be the power consumption price per unit of net
mining rate and per unit of time. It is easy to see that the power consumption
costs per unit of time with respect to the honest and dishonest mining pools
are given by $c_{P}\left(  \alpha_{1}-\gamma\right)  $ and $c_{P}\left(
\widetilde{\alpha}_{2}+\gamma\right)  $, respectively. (b) Administrative
cost. Let $c_{A}$ be the administrative price per unit of real mining rate and
per unit of time. Then the administrative costs per unit of time with respect
to the honest and dishonest mining pools are given by $c_{A}\left(  \alpha
_{1}-\gamma\right)  $ and $c_{A}\left(  \widetilde{\alpha}_{2}+\gamma\right)
\left(  1+\tau\right)  $, respectively.

\textbf{Competitive criterions:} In the blockchain selfish mining, the honest
and dishonest mining pools compete fiercely in finding the nonces to generate the blocks, and they publish the blocks to make two block branches forked at a common tree root. For the two block branches, the longer block branch in
the forked structure is called a \emph{main chain}, which or the part of which will be pegged
on the blockchain. Under the selfish mining attacks, such two mining pools follow the different competitive criterions.

(a) A two-block leading competitive criterion for the honest mining pool. The honest chain of blocks is taken as the main chain pegged on the blockchain, as soon as the honest chain of blocks is two blocks ahead of the dishonest chain of blocks.

(b) A modification of two-block leading competitive criterion for the
dishonest mining pool. Once the
dishonest chain of blocks is two blocks ahead of the honest chain of blocks,
the dishonest chain of blocks can be taken as the main chain. To get
more reward, the dishonest mining pool may prefer to
keep its mined blocks secret, and continue to mine more blocks rather than broadcast all the
mined information.

Since the dishonest miners are minority, their mining power is limited, the
dishonest mining pool will not be extend infinitely. We assume that once the
dishonest main chain contains $m$ blocks, its part $n$ blocks $(n\leq m)$ must be
pegged on the blockchain immediately. In addition, the limitation of the dishonest
main chain leads to that the honest main chain containing at most $n-2$ blocks due to
the two-block leading competitive criterion.

\textbf{Blockchain-pegged processes: }If the main chain is formed, then the
mining processes are terminated immediately. The honest main chain or the part of the dishonest main chain is pegged on the blockchain, and the
blockchain-pegged times are i.i.d. and exponential with mean $1/\mu.$ The mining pool
of the main chain can obtain an appropriate amount of reward (or
compensation) from two different parts: A block reward $r_{B}$ by the
blockchain system and an average total transaction fee $r_{F}$ in the block. At the
same time, all the blocks of the other non-main chain become orphan and immediately return to the transaction pool without any new fee. Note that no new blocks
are generated during the blockchain-pegged process of the main chain.

\begin{figure}[htb]
%\centering
\includegraphics[width=12cm]{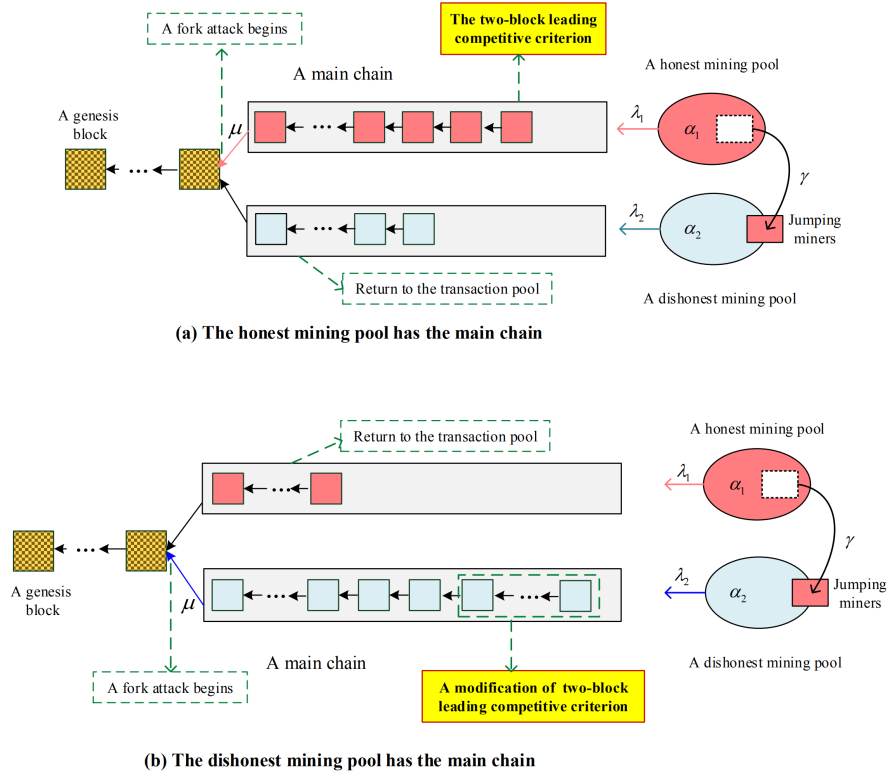}  \caption{A blockchain
selfish mining with two different mining pools.}%
\label{fig1}%
\end{figure}

We assume that all the random variables defined above are independent of each other. Fig.~\ref{fig1} provides an intuitive understanding for the two cases.
\section{Optimization Model Formulation}

In this section, we establish an optimization problem to find an optimal
blockchain-pegged policy for the dishonest mining pool. To do this, we set up
a policy-based continuous-time Markov process and
introduce some key factors.
\subsection{The states and policies}

To study the blockchain-pegged policy of the blockchain selfish mining with two different mining pools, we
first define both `states' and `policies' to express such a stochastic dynamic.

Let $N_{1}(t)$ and $N_{2}(t)$ be the numbers of blocks mined by the honest and
dishonest mining pools at time $t$, respectively. Then $\left(  N_{1}%
(t),N_{2}(t)\right)  $ is regarded as the state of this system at time $t$.
Obviously, all the cases of State $\left(  N_{1}(t),N_{2}(t)\right)  $ form a
state space as follows:%
\[
\bm       \Omega=\underset{k=0}{\overset{m+2}{\bigcup}}{{{\bm       \Omega
_{k},}}}%
\]
where%
\begin{align*}
{{{\bm  \Omega_{0}}}}  &  {=}{{}}\left\{  \left(  0,0\right)  ,\left(
0,1\right)  ,\ldots,\left(  0,m\right)  \right\}  ,\\
{{{\bm  \Omega_{1}}}}  &  {=}{{}}\left\{  \left(  1,0\right)  ,\left(
1,1\right)  ,\ldots,\left(  1,m\right)  \right\}  ,\\
{{{\bm  \Omega}}}_{k}  &  {=}{{}}\left\{  \left(  k,k-2\right)  ,\left(
k,k-1\right)  ,\ldots,\left(  k,m\right)  \right\}  ,k=2,3,\ldots,m+2.
\end{align*}

Actually, the blockchain-pegged policy of the dishonest mining pool can be represented by blockchain-pegged probability $p$. The dishonest mining pool pegs the
main chain on the blockchain according to the probability $p$ at the state $\left(n_{1},n_{2}\right)  $ for $\left(n_{1},n_{2}\right)  \in\bm                \Omega$.
From the problem description in Section 2, it is easy to see that%
\begin{equation}
p=\left\{
\begin{array}
[c]{cl}%
a\in\left[  0,1\right]  , & n_{1}=0,1,\ldots,m-3,\text{ }n_{2}=n_{1}%
+2,n_{1}+3,\ldots,m-1,\\
1, & n_{1}=0,1,\ldots,m-2,\text{ }n_{2}=m,\\
0, & \text{otherwise.}%
\end{array}
\right.  \label{1}%
\end{equation}
It is obviously that the Markov process is controlled by the blockchain-pegged policy (the probability $p$). Let all the possible
probabilities $p$ given in (\ref{1}) compose a policy space as follows:%
\[
\mathcal{P}=\left\{  p:p\in\left[  0,1\right]  ,\text{ for }\left(
n_{1},n_{2}\right)  \in\bm                \Omega\right\}  .
\]

\begin{figure}[htp]
%\centering
\includegraphics[width=\textwidth]{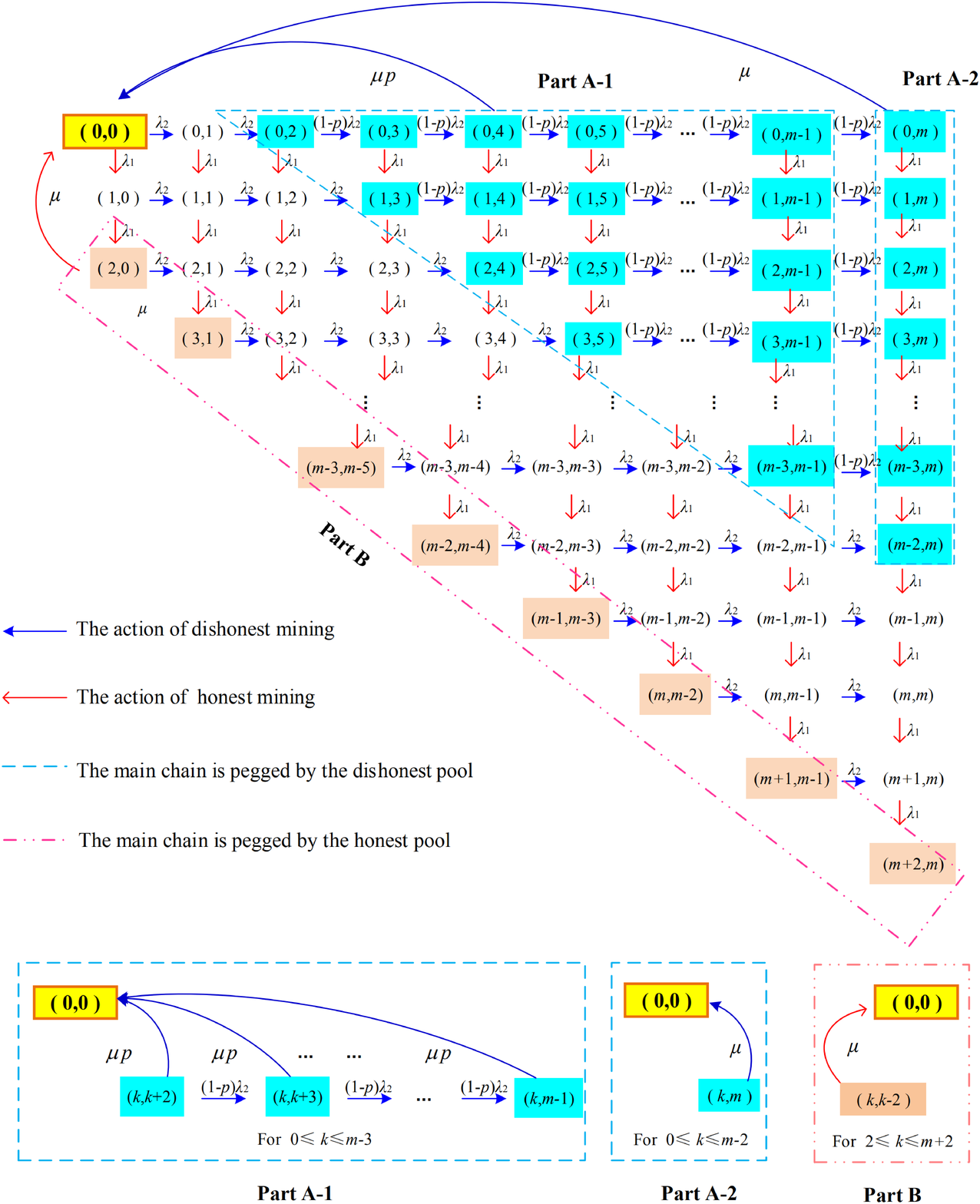}
\caption{The state transition relation of the Markov process.}%
\label{fig2}%
\end{figure}

It is readily seen that State $(0,0)$ is a key state, which plays a key role in setting up the Markov process of
two block branches forked at the tree root. In fact, State $(0,0)$ describes the
tree root as the starting point of the fork attacks, e.g., see Fig.~\ref{fig2}. If the
Markov process enters State $(0,0)$, then the fork attack ends immediately,
and the main chain is pegged on the blockchain.

Now, from Fig.~\ref{fig2}, we provide an interpretation for the
blockchain-pegged probability $p$ as follows:

(1) In Part A-1, i.e., $n_{1}=0,1,\ldots,m-3\ $and $n_{2}=n_{1}+2,n_{1}%
+3,\ldots,m-1,$ the dishonest mining pool follows the modification of
two-block leading competitive criterion and forms the dishonest main chain,
then the probability $p\in\left[  0,1\right]  $.

(2) In Part A-2, i.e., $n_{1}=0,1,\ldots,m-2\ $and $n_{2}=m,$ for the
limitation of dishonest mining power, the dishonest main chain must be pegged
on the blockchain, or there is a risk of getting no reward. It is easy to see
that the probability $p$ is taken as 1.

(3) In the rest of Fig.~\ref{fig2}, it is the competitive process of honest and dishonest mining pools. Therefore, $p=0$ for the
dishonest main chain hasn't formed. In addition, the states in Part B mean that
the honest main chain is formed.

Due to the modification of two-block leading competitive criterion, the limitation $m$ of the dishonest mining pool must be
more than 2, so that there exist the blockchain-pegged policy for the
dishonest mining pool. If
$m\geq5,$ the infinitesimal generator has a general expression (note that the special cases of $m=3$ and $m=4$ are omitted here). In what
follows, we assume $m\geq5$ for convenience of calculation, but the analysis
method is similar.

Let $\mathbf{X}^{(p)}\left(  t\right)  =\left(  N_{1}(t),N_{2}(t)\right)
^{(p)}\ $be the system state at time $t$ under any given policy $p\in
\mathcal{P}$. Then $\left\{  \mathbf{X}^{(p)}\left(  t\right)  :t\geq
0\right\}  $ is a policy-based continuous-time Markov process on the state
space $\bm                                                      \Omega$ whose
state transition relation is depicted in Fig.~\ref{fig2}. Obviously, such a Markov
process is a special form of the pyramid Markov process given in Li et al. \cite{Li:2020}.
Based on this, the infinitesimal generator of the Markov process $\left\{  \mathbf{X}^{(p)}\left(  t\right) \! :\!t\geq
0\right\}  $ is given by%
\begin{equation}
\mathbf{Q}^{(p)}=\left(
\begin{array}
[c]{cccccc}%
Q_{0,0} & B_{0} &  &  &  & \\
Q_{1,0} & Q_{1,1} & B_{1} &  &  & \\
Q_{2,0} &  & Q_{2,2} & B_{2} &  & \\
\vdots &  &  & \ddots & \ddots & \\
Q_{m+1,0} &  &  &  & Q_{m+1,m+1} & B_{m+1}\\
Q_{m+2,0} &  &  &  &  & Q_{m+2,m+2}%
\end{array}
\right)  . \label{5}%
\end{equation}
Here, we omit the details of the submatrices in the infinitesimal generator $\mathbf{Q}^{(p)}$.
\subsection{The stationary probability vector}

Based on some special properties of the infinitesimal generator, we provide
the stationary probability vector for the policy-based continuous-time Markov
process $\left\{  \mathbf{X}^{(p)}\left(  t\right)  :t\geq0\right\}  $.

For $n_{1}=0,1,\ldots,m-3$, $n_{2}=n_{1}+2,n_{1}+3,\ldots,m-1$ and $0\leq
p<1,$ it is clear from the finite states that the policy-based continuous-time Markov process
$\mathbf{Q}^{(p)}$ must be irreducible, aperiodic and positive recurrent.

We write the stationary probability vector of the Markov process $\left\{
\mathbf{X}^{(p)}\left(  t\right)  :t\!\geq\!0\right\}  $\ as follows:%
\begin{equation}
\bm \pi^{(p)}=\left(  \bm \pi_{0}^{(p)},\bm \pi_{1}^{(p)},\ldots,\bm \pi
_{m+2}^{(p)}\right)  ,\label{3}%
\end{equation}
where%
\begin{align*}
\bm \pi_{0}^{(p)} &  =\left(  \pi^{(p)}\left(  0,0\right)  ,\pi^{(p)}\left(
0,1\right)  ,\ldots,\pi^{(p)}\left(  0,m\right)  \right)  ,\\
\bm \pi_{1}^{(p)} &  =\left(  \pi^{(p)}\left(  1,0\right)  ,\pi^{(p)}\left(
1,1\right)  ,\ldots,\pi^{(p)}\left(  1,m\right)  \right)  ,\\
\bm \pi_{k}^{(p)} &  =\left(  \pi^{(p)}\left(  k,k-2\right)  ,\pi^{(p)}\left(
k,k-1\right)  ,\ldots,\pi^{(p)}\left(  k,m\right)  \right)  ,\text{ }2\leq
k\leq m+2.
\end{align*}

Let
\begin{align}
\mathbf{D}_{0}  &  =1,\nonumber\\
\mathbf{D}_{k}  &  =B_{k-1}\left(  -Q_{k,k}\right)  ^{-1},\text{ }%
k=1,2,\ldots,m+2. \label{23-1}%
\end{align}
Then the following theorem provides an explicit expression for the stationary
probability vector $\bm        \pi^{(p)}$ by means of the system of linear equations: $\bm             \pi^{(p)}\bm    Q^{(p)}%
=\mathbf{0}$ and $\bm             \pi
^{(p)}\bm     e=1.$%.
\begin{theorem}
The stationary probability vector $\bm                         \pi^{(p)}$ of
the Markov process $\bm    Q^{(p)}$ is given by%
\begin{equation}
\bm  \pi_{k}^{(p)}=\bm  \pi_{0}^{(p)}\underset{l=1}{\overset{k}{\prod}%
}\mathbf{D}_{l},\text{ } \label{6}%
\end{equation}
where $\bm                                                        \pi
_{0}^{(p)}$\ is determined by the system of linear equations%
\begin{align*}
\bm  \pi_{0}^{(p)}\left(  \underset{k=0}{\overset{m+2}{\sum}}\underset
{l=0}{\overset{k}{\prod}}\mathbf{D}_{l}Q_{k,0}\right)   &  =\mathbf{0},\\
\bm  \pi_{0}^{(p)}\left(  \underset{k=0}{\overset{m+2}{\sum}}\underset
{l=0}{\overset{k}{\prod}}\mathbf{D}_{l}\bm  e\right)   &  =1.
\end{align*}
\end{theorem}
\subsection{The reward function}

A reward function of the dishonest mining pool with respect to
both states and policies is defined as a profit rate (i.e., the total revenues
minus the total costs per unit of time).

Let $R=r_{B}+r_{F}$ and $C=\left(  \widetilde{\alpha}+\gamma\right)  \left[
c_{P}+c_{A}\left(  1+\tau\right)  \right]  $. Then $R$ and $C$ denote the
blockchain-pegged reward and the mining cost for the dishonest mining pool,
respectively. According to Fig.~\ref{fig2}, the reward function at State $\left(
N_{1}\left(  t\right)  ,N_{2}\left(  t\right)  \right)  ^{(p)}$ under the blockchain-pegged
policy $p$ is defined as follows:
\[
f^{(p)}\left(  n_{1},n_{2}\right)  =\left\{
\begin{array}
[c]{ll}%
n_{2}R\mu p-C, & \text{if }0\leq n_{1}\leq m-3\text{ and }n_{1}+2\leq
n_{2}\leq m-1\text{,}\\
mR\mu-C, & \text{if }0\leq n_{1}\leq m-2\text{ and }n_{2}=m\text{,}\\
-C, & \text{otherwise.}%
\end{array}
\right.
\]
We futher define a column vector $\bm f^{(p)}$ composed of the elements
$f^{(p)}\left(  n_{1},n_{2}\right)  $ as%
\begin{equation}
\bm f^{(p)}=\left(  \left(  \bm f_{0}^{(p)}\right)  ^{T},\left(  \bm
f_{1}^{(p)}\right)  ^{T},\ldots,\left(  \bm f_{m+2}^{(p)}\right)  ^{T}\right)
^{T},\label{8}%
\end{equation}
where%
\begin{align*}
\bm f_{0}^{(p)} &  =\left(  f^{(p)}\left(  0,0\right)  ,f^{(p)}\left(
0,1\right)  ,\ldots,f^{(p)}\left(  0,m\right)  \right)  ^{T},\\
\bm f_{1}^{(p)} &  =\left(  f^{(p)}\left(  1,0\right)  ,f^{(p)}\left(
1,1\right)  ,\ldots,f^{(p)}\left(  1,m\right)  \right)  ^{T},\\
\bm f_{k}^{(p)} &  =\left(  f^{(p)}\left(  k,k-2\right)  ,f^{(p)}\left(
k,k-1\right)  ,\ldots,f^{(p)}\left(  k,m\right)  \right)  ^{T}\text{,
}k=2,3,\ldots,m+2.
\end{align*}

In the remainder of this section, the long-run average profit of the dishonest
mining pool under a blockchain-pegged policy $p$ is defined as%
\begin{align}
\eta^{p}  =\lim_{T\rightarrow+\infty}E\left\{  \frac{1}{T}\int_{0}%
^{T}f^{(p)}\left(  \left(  N_{1}\left(  t\right)  ,N_{2}\left(  t\right)
\right)  ^{(p)}\right)  \text{d}t\right\} =\bm  \pi^{\left(  p\right)  }\bm  f^{\left(  p\right)  }, \label{9}%
\end{align}
where $\bm      \pi^{\left(  p\right)  }$ and $\bm      f^{\left(  p\right)
}$ are given by (\ref{6}) and (\ref{8}), respectively.
\subsection{The performance potential}

The sensitivity-based optimization theory has a fundamental quantity called performance potential by Cao
\cite{Cao:2007}, which is defined as
\begin{equation}
g^{\left(  p\right)  }\left(  n_{1},n_{2}\right)  =E\left\{  \left.  \int
_{0}^{+\infty}\left[  f^{(p)}\left(  \mathbf{X}^{\left(  p\right)  }\left(
t\right)  \right)  -\eta^{p}\right]  \text{d}t\right|  \mathbf{X}^{\left(
p\right)  }\left(  0\right)  =\left(  n_{1},n_{2}\right)  \right\}  ,
\label{11}%
\end{equation}
where $\eta^{p}$ is defined in (\ref{9}). For any blockchain-pegged policy $p\in\mathcal{P}$, $g^{\left(
p\right)  }\left(  n_{1},n_{2}\right)  $ quantifies the contribution of the
initial State $\left(  n_{1},n_{2}\right)  $ to the long-run average profit of
the dishonest mining pool. Here, $g^{\left(  p\right)  }\left(  n_{1}%
,n_{2}\right)  $ is also called the relative value function or the bias in the
traditional MDP theory, see, e.g., Puterman \cite{Put:1994}. We further define
a column vector $\bm      g^{\left(  p\right)  }$ as%
\begin{equation}
\bm  g^{(p)}=\left(  \left(  \bm  g_{0}^{(p)}\right)  ^{T},\left(  \bm
g_{1}^{(p)}\right)  ^{T},\ldots,\left(  \bm  g_{m+2}^{(p)}\right)
^{T}\right)  ^{T}, \label{12}%
\end{equation}
where%
\begin{align*}
\bm  g_{0}^{(p)}  &  =\left(  g^{(p)}\left(  0,0\right)  ,g^{(p)}\left(
0,1\right)  ,\ldots,g^{(p)}\left(  0,m\right)  \right)  ^{T},\\
\bm  g_{1}^{(p)}  &  =\left(  g^{(p)}\left(  1,0\right)  ,g^{(p)}\left(
1,1\right)  ,\ldots,g^{(p)}\left(  1,m\right)  \right)  ^{T},\\
\bm  g_{k}^{(p)}  &  =\left(  g^{(p)}\left(  k,k-2\right)  ,g^{(p)}\left(
k,k-1\right)  ,\ldots,g^{(p)}\left(  k,m\right)  \right)  ^{T},\text{
}k=2,3,\ldots,m+2.
\end{align*}

A similar computation to that in Ma et al. \cite{Ma:2019a, Ma:2019b} is omitted here, we can provide an expression for the vector ${\bm
g}^{(p)}$
\begin{equation}
{\bm  g}^{(p)}=R\bm  a+\bm  b, \label{g-1}%
\end{equation}
where $\bm  a$ and $\bm  b$ can be given by $\mathbf{Q}^{\left(  p\right)}$, $\bm  \pi^{(p)}$ and  $\bm f^{(p)}$. It is seen that all the entries $g^{\left(  p\right)  }\left(  n_{1}%
,n_{2}\right)  \ $in ${\bm      g}^{(p)}$ are the linear functions
of $R$. Therefore, our objective is to find the optimal blockchain-pegged policy $p^{\ast}$ such that the long-run average profit of the dishonest
mining pool $\eta^{p}$ is maximize, that is,%
\begin{equation}
p^{\ast}=\underset{p\in\mathcal{P}}{\arg\max}\left\{  \eta^{p}\right\}  .
\label{10}%
\end{equation}

However, it is very challenging to analyze some interesting
structure properties of the optimal blockchain-pegged policy $p^{\ast}.$
In the remainder of this paper, we will apply the
sensitivity-based optimization theory to study such an optimal problem.
\section{Monotonicity and Optimality}

In this section, we use the the
sensitivity-based optimization theory to discuss monotonicity and optimality of
the long-run average profit of the dishonest mining pool with respect to the
blockchain-pegged policy. Based on this, we obtain the optimal
blockchain-pegged policy of the dishonest mining pool.

In an MDP, system policies will affect the element values of infinitesimal generator and reward function. That is, if the policy $p$ changes, then the infinitesimal generator
$\mathbf{Q}^{\left(  p\right)  }$ and the reward function $\bm   f^{\left(
p\right)  }$ will have their corresponding changes. To express such a change
mathematically, we take two different policies
$p,p^{\prime}\in\mathcal{P}$, both of which correspond to their infinitesimal
generators $\mathbf{Q}^{\left(  p\right)  }$ and $\mathbf{Q}^{\left(
p^{\prime}\right)  }$, and to their reward functions $\bm    f^{\left(
p\right)  }$ and $\bm               f^{\left(  p^{\prime}\right)  }$.

The following lemma provides the performance difference equation for the difference
$\eta^{p^{\prime}}-\eta^{p}$ of the long-run average performances for any two blockchain-pegged policies
$p,p^{\prime}\in\mathcal{P}$. Here, we only restate it without proof, while
readers may refer to Cao \cite{Cao:2007} and Ma et al. \cite{Ma:2019a} for more details.
\begin{lemma}
\label{Lem:diff}For any two blockchain-pegged policies $p,p^{\prime}%
\in\mathcal{P}$, we have
\begin{equation}
\eta^{p^{\prime}}-\eta^{p}=\bm  \pi^{\left(  p^{\prime}\right)  }\left[
\left(  \mathbf{Q}^{\left(  p^{\prime}\right)  }-\mathbf{Q}^{\left(  p\right)
}\right)  \bm  g^{(p)}\mathbf{+}\left(  \bm  f^{\left(  p^{\prime}\right)
}-\bm  f^{\left(  p\right)  }\right)  \right]  . \label{30}%
\end{equation}
\end{lemma}

Therefore, to find the optimal blockchain-pegged policy $p^{\ast},$ we
consider such two blockchain-pegged policies $p,p^{\prime}\in\mathcal{P}%
$. Suppose the blockchain-pegged policy is changed from $p$ to
$p^{\prime},$ which corresponding the states $\left(  n_{1},n_{2}\right)  $ for $n_{1}=0,1,\ldots,m-3$ and $n_{2}=n_{1}%
+2,n_{1}+3,\ldots,m-1$, i.e., Part A-1 of Fig. 2.

Using Lemma 2, we examine the sensitivity of blockchain-pegged policy on
the long-run average profit of the dishonest mining pool. Substituting
(\ref{5}) and (\ref{8}) into (\ref{30}), we have %
\begin{align}
&  \text{ \ \ \ }\eta^{p^{\prime}}-\eta^{p}\nonumber\\
&  =\bm  \pi^{\left(  p^{\prime}\right)  }\left[  \left(  \mathbf{Q}^{\left(
p^{\prime}\right)  }-\mathbf{Q}^{\left(  p\right)  }\right)  \bm
g^{(p)}\mathbf{+}\left(  \bm  f^{\left(  p^{\prime}\right)  }-\bm  f^{\left(
p\right)  }\right)  \right] \nonumber\\
&  =\left(  p^{\prime}\!-\!p\right)  \underset{n_{1}=0}{\overset{m-3}{\sum}%
}\underset{n_{2}\!=\!n_{1}\!+\!2}{\overset{m-1}{\sum}}\pi^{\left(  p^{\prime}\right)
}\left(  n_{1},n_{2}\right)  \left[  \mu\!-\!\left(  \mu\!-\!\lambda_{2}\right)
g^{\left(  p\right)  }\left(  n_{1},n_{2}\right) \! -\!\lambda_{2}g^{\left(
p\right)  }\left(  n_{1},n_{2}\!+\!1\right) \! +\!n_{2}R\mu\right]  .\label{34}
\end{align}

With the difference (\ref{34}), we can easily obtain the following equation%
\begin{equation}
\frac{\triangle\eta^{p}}{\triangle p}\!=\!\underset{n_{1}=0}{\overset{m-3}{\sum}%
}\underset{n_{2}=n_{1}\!+\!2}{\overset{m-1}{\sum}}\pi^{\left(  p^{\prime}\right)
}\left(  n_{1},n_{2}\right)  \left[  \mu\!-\!\left(  \mu-\lambda_{2}\right)
g^{\left(  p\right)  }\left(  n_{1},n_{2}\right) \! -\!\lambda_{2}g^{\left(
p\right)  }\left(  n_{1},n_{2}\!+\!1\right)  \!+\!n_{2}R\mu\right]  , %
\end{equation}
where $\triangle\eta^{p}=\eta^{p^{\prime}}-\eta^{p}$ and $\triangle
p=p^{\prime}-p$. As $p^{\prime}\rightarrow p$,
\[
\left.  \frac{\text{d}\eta^{p}}{\text{d}p}\right|  _{\triangle p\rightarrow
0}=\underset{\triangle p\rightarrow0}{\lim}\frac{\eta^{p^{\prime}}-\eta^{p}%
}{\triangle p},
\]
we derive the following derivative equation%
\begin{equation}
\frac{\text{d}\eta^{p}}{\text{d}p}\!=\!\underset{n_{1}=0}{\overset{m-3}{\sum}%
}\underset{n_{2}=n_{1}+2}{\overset{m-1}{\sum}}\pi^{\left(  p\right)  }\left(
n_{1},n_{2}\right)  \left[  \mu\!-\!\left(  \mu-\lambda_{2}\right)  g^{\left(
p\right)  }\left(  n_{1},n_{2}\right) \! -\!\lambda_{2}g^{\left(  p\right)
}\left(  n_{1},n_{2}\!+\!1\right) \! +\!n_{2}R\mu\right]  . \label{37-1}%
\end{equation}
According to (\ref{g-1}), $g^{\left(  p\right)  }\left(  n_{1},n_{2}\right)  $
and $g^{\left(  p\right)  }\left(  n_{1},n_{2}+1\right)  $\ are both linear
functions w.r.t. $R.$ Thus, we denote\ $g^{\left(  p\right)  }\left(
n_{1},n_{2}\right)  $ and $g^{\left(  p\right)  }\left(  n_{1},n_{2}+1\right)
$ as $a_{n_{1},n_{2}}R+b_{n_{1},n_{2}}$ and$\ a_{n_{1},n_{2}+1}R+b_{n_{1}%
,n_{2}+1},\ $respectively. Substituting into
(\ref{37-1}), we have
\begin{equation}
\frac{\text{d}\eta^{p}}{\text{d}p}=\overline{a}R+\overline{b}, \label{40}%
\end{equation}
where%
\begin{align*}
\overline{a}  &  =\underset{n_{1}=0}{\overset{m-3}{\sum}}\underset{n_{2}%
=n_{1}+2}{\overset{m-1}{\sum}}\pi^{\left(  p\right)  }\left(  n_{1}%
,n_{2}\right)  \left[  \left(  \lambda_{2}-\mu\right)  a_{n_{1},n_{2}}%
-\lambda_{2}a_{n_{1},n_{2}+1}+n_{2}\mu\right]  ,\\
\overline{b}  &  =\underset{n_{1}=0}{\overset{m-3}{\sum}}\underset{n_{2}%
=n_{1}+2}{\overset{m-1}{\sum}}\pi^{\left(  p\right)  }\left(  n_{1}%
,n_{2}\right)  \left[  \left(  \lambda_{2}-\mu\right)  b_{n_{1},n_{2}}%
+\lambda_{2}a_{n_{1},n_{2}+1}b_{n_{1},n_{2}+1}+\mu\right]  .
\end{align*}
It is clear that $\frac{\text{d}\eta^{p}}{\text{d}p}$ is also a linear
function w.r.t. $R,$ and depends only on the current policy.

\begin{Rem}
It is seen from (16) that we only need to know the sign of $\frac{\text{d}\eta^{p}}{\text{d}p}$, instead of its precise value. The estimation accuracy of a sign is usually better than that of a value. Therefore, this feature can help us find the optimal blockchain-pegged policy effectively. Moreover, we see that we do not have to know some prior system information. Thus, the complete system information is not required in our approach and this is an advantage during the practical application.
\end{Rem}

\begin{Rem}
The key idea of the sensitivity-based optimization theory is to utilize the performance sensitivity information, such as the performance difference, to conduct the optimization of stochastic systems. Therefore, even if the competition criteria become more complicated, it does not affect the applicability of our method.
\end{Rem}

The following theorems discuss monotonicity and optimality of the long-run
average profit $\eta^{p}$ of the dishonest mining pool with respect to the
blockchain-pegged policy $p.$

\begin{theorem}
If $R>-\overline{b}/\overline{a}$,\ then the long run average profit $\eta
^{p}$ is strictly monotone increasing with respect to each decision element
$p\in\left[  0,1\right]  $, and the optimal blockchain-pegged policy $p^{\ast}=1.$
\end{theorem}

This theorem follows directly (\ref{40}). It is seen that the optimal
blockchain-pegged policy $p^{\ast}=1$ just corresponding to any State
$\left(  n_{1},n_{1}+2\right)  \ $in Part A-1 of Fig. 2, and the state transition has changed. In this
case, the dishonest chain of blocks is only two blocks ahead of the honest
chain of blocks, the dishonest mining pool should peg on the blockchain, also follows the
two-block leading competitive criterion.

Therefore, when the blockchain-pegged reward is higher with $R>-\overline
{b}/\overline{a},$ it is seen that the dishonest miners become honest, all miners will follow the PoW protocol and broadcast to the whole community. In this case, the selfish mining attacks should be invalid.

\begin{theorem}
If $0\leq R<-\overline{b}/\overline{a}$, then the long run average profit
$\eta^{p}$ is strictly monotone decreasing with respect to each decision
element $p\in\left[  0,1\right]  $, and the optimal blockchain-pegged
policy $p^{\ast}=0.$
\end{theorem}

Simlar to Theorem 2, this theorem also follows directly (\ref{40}). It is seen
that the optimal blockchain-pegged policy $p^{\ast}=0$ corresponding to
any State $\left(  n_{1},n_{2}\right)  \ $in Part A-1 of Figure 2$\mathbf{.}$

In the blockchain selfish mining, if the dishonest mining pool makes decision
not to peg on the blockchain, i.e., $p^{\ast}=0,$ the main chain is detained
to continue mining more blocks so that it is not broadcasted in the blockchain
network, until the number of blocks reaches $m$ for the limited mining bound. In this case, the
dishonest mining pool prefer to obtain more mining profit through winning on
mining more blocks, rather than peg on the blockchain prematurely.

Therefore, when the blockchain-pegged reward is lower with $0\leq R<-\overline{b}/\overline{a}$, it is seen that the dishonest mining pool
follows the $m$-block leading competitive criterion under the selfish mining
attacks.
\begin{theorem}
If $R=-\overline{b}/\overline{a}$,\ then the change of blockchain-pegged
policy $p$ no longer improve the long-run average profit $\eta^{p}.$
\end{theorem}

With Theorem 4, the dishonest miners don't care about when the main chain is
pegged on the blockchain, thus the blockchain-pegged policy can be chosen
randomly in set $\left[  0,1\right]  .$

\section{Concluding Remarks}

In this paper, we propose a novel dynamic decision method by applying the
sensitivity-based optimization theory to study the optimal blockchain-pegged
policy of blockchain selfish mining with two different mining pools.

We describe a more general blockchain selfish mining with a modification of two-block leading competitive
criterion, which is related to the blockchain-pegged policies. To
find the optimal blockchain-pegged policy of the dishonest mining pool,
we analyze the monotonicity and optimality
of the long-run average profit with respect to the blockchain-pegged
policy under some restrained blockchain-pegged rewards. We prove the
structure of optimal blockchain-pegged policy with respect to the
blockchain-pegged rewards. Different from those previous works in the
literature on applying the traditional MDP theory to the blockchain selfish
mining, the sensitivity-based optimization theory used in this paper is easier and
more convenient in the optimal policy study of blockchain selfish mining.

Along such a research line of applying the sensitivity-based optimization
theory, there are a number of interesting directions for potential future
research, for example:

$\bullet$ Extending to the blockchain selfish mining with multiple mining
pools, for example, a different competitive criterion, no space limitation of the dishonest pool
and so on;

$\bullet$ analyzing non-Poisson inputs such as Markovian arrival processes
(MAPs) and/or non-exponential service times, e.g. the PH distributions;

$\bullet$ discussing the long-run average performance is influenced by some
concave or convex reward (or cost) functions; and

$\bullet$ studying individual or social optimization for the blockchain
selfish mining from a perspective of combining game theory with the
sensitivity-based optimization.
%
% ---- Bibliography ----
%
% BibTeX users should specify bibliography style 'splncs04'.
% References will then be sorted and formatted in the correct style.
%
% \bibliographystyle{splncs04}
% \bibliography{mybibliography}
%

\end{document}